\documentclass[reprint,superscriptaddress,amsmath,amssymb,aps,prr,longbibliography]{revtex4-1}
\usepackage{graphicx}
\usepackage{dcolumn}
\usepackage{bm}
\usepackage{amsmath}
\usepackage{braket}
\usepackage{physics}
\usepackage{amsfonts}
\usepackage{amssymb}
\newcommand{\nn}{\nonumber}

\newcommand{\uz}{{\hat{u}}^z}
\newcommand{\p}{\ket{\Psi_{u}}}
\begin{document}
\preprint{APS/123-QED}
\title{Topological Nematic Phase Transition 
\\ in Kitaev Magnets Under Applied Magnetic Fields}
\author{Masahiro O. Takahashi}
\email{takahashi@blade.mp.es.osaka-u.ac.jp}
\affiliation{Department of Materials Engineering Science, Osaka University, Toyonaka 560-8531, Japan}
\author{Masahiko G. Yamada}
\affiliation{Department of Materials Engineering Science, Osaka University, Toyonaka 560-8531, Japan}
\author{Daichi Takikawa}
\affiliation{Department of Materials Engineering Science, Osaka University, Toyonaka 560-8531, Japan}
\author{Takeshi Mizushima}
\affiliation{Department of Materials Engineering Science, Osaka University, Toyonaka 560-8531, Japan}
\author{Satoshi Fujimoto}
\affiliation{Department of Materials Engineering Science, Osaka University, Toyonaka 560-8531, Japan}
\affiliation{Center for Quantum Information and Quantum Biology, Osaka University, Toyonaka 560-8531, Japan}
\date{\today}
\begin{abstract}
We propose a scenario of realizing the toric code phase, which can be potentially utilized for fault-tolerant quantum computation,
 in candidate materials of Kitaev magnets. 
 It is demonstrated that four-body interactions among Majorana fermions in the Kitaev spin liquid state,
 which are induced by applied magnetic fields as well as non-Kitaev-type exchange interactions, trigger
 a nematic phase transition of Majorana bonds without magnetic orders, 
 accompanying the change of the Chern number from $\pm 1$ to zero.
 This gapful spin liquid state with zero Chern number is nothing but the toric code phase.
Our result potentially explains the topological nematic transition recently observed in
$\alpha$-RuCl$_3$ via heat capacity measurements [O. Tanaka \textit{et al.}, arXiv:2007.06757].
\end{abstract}

\maketitle

\section{Introduction}
A quantum spin liquid (QSL) is an exotic phase of matter, 
where spins do not have long-range order even at zero temperature.
The Kitaev model~\cite{Kitaev2006} is an exactly solvable model of QSLs, which is
defined on the honeycomb lattice with bond-dependent Ising interactions.
A remarkable feature of the Kitaev model is that the system is
described by Majorana fermions interacting with $Z_2$ gauge fields, allowing the realization of Abelian and non-Abelian anyons.
For almost isotropic bond-dependent Ising interactions,
the systems exhibits a chiral QSL phase with non-Abelian anyons, 
when an applied magnetic field opens an energy gap of Majorana fermions.
On the other hand, for highly anisotropic Ising interactions, the toric code phase with Abelian anyons is realized.
It is noted that both of these phases can be utilized for topological quantum computation~\cite{Kitaev2006,KITAEV20032}.
After the Kitaev's seminal paper, various phenomena associated with Majorana fermions in Kitaev magnets have been extensively explored~\cite{PhysRevLett.98.247201,PhysRevLett.98.087204,Knolle2014,Nasu2015,Hermanns2015WSL,Hermanns2015,Obrien2016,Song2016,Nasu2017,Udagawa2018,Yamada2017XSL,Yamada2020,Takikawa2019,Takikawa2020}. 
Since Jackeli and Khaliullin~\cite{Jackeli2009,PhysRevLett.113.107201,Rau2014,PhysRevB.93.155143,Winter2016,Yamada2017MOF} pointed out that
isotropic bond-dependent Ising interactions arise in some honeycomb layered materials,
the experimental search for candidate materials has been an important subject.
Several materials including $4d^5$ or $5d^5$ transition metal ions
with a strong spin-orbit coupling have been proposed for Kitaev magnets,
such as $\mathrm{Na_2IrO_3}$~\cite{Lee2010,Singh2010}, $\alpha$-$\mathrm{Li_2IrO_3}$~\cite{Singh2012}, and $\alpha$-RuCl$_3$~\cite{Plumb2014}.
$\alpha$-RuCl$_3$ under an applied magnetic field is one of the best
candidates for the Kitaev magnet~\cite{Baek2017}.
For this material, a signature of Majorana fermions in the chiral QSL phase is observed 
via the measurement of the half-integer thermal quantum Hall effect~\cite{Kasahara2018,Vinkler2018,Ye2018}.
Moreover, according to the recent specific heat measurements~\cite{Tanaka2020},
approximated threefold rotational symmetry of the honeycomb lattice is drastically broken to obviously twofold one at high fields, implying that
nematic-type phase transition occurs~\footnote{Although the crystal structure of  $\alpha$-RuCl$_3$ may be monoclinic rather than trigonal~\cite{PhysRevB.92.235119,PhysRevB.93.155143}, the field angle dependence of heat capacity observed in $\alpha$-RuCl$_3$ by Tanaka et al.~\cite{Tanaka2020} shows the $C_3$ symmetry within error bars for magnetic fields $< 9$ T. This approximated threefold rotational symmetry drastically breaks down to obvious twofold one for strong magnetic fields $> 9$ T, which implies the possibility of nematic phase transition in $\alpha$-RuCl$_3$~\cite{Tanaka2020}}.
 Remarkably, the nematic phase transition accompanies the disappearance of
the half-quantized thermal Hall effect~\cite{Kasahara2018}, though experimental results suggest that the system may be still in a QSL phase.
To understand these observations, we need another ingredient missing in previous research for Kitaev magnets,
which includes various additional spin-spin interactions, such as Heisenberg, $\Gamma$, and $\Gamma^{\prime}$ terms.

In this work, we theoretically discuss the possibility of a nematic phase transition
due to many-body interactions among itinerant Majorana fermions in Kitaev magnets.
Effects of interactions between Majorana fermions have been extensively studied
for Kitaev magnets as well as topological superconductors~\cite{Stoudenmire2011,Niu2012,Thomale2013,Sticlet2014,Hermanns2015,Rahmani20152,Affleck2017,Li2018,Wamer2018,Rahmani2019,Rahmani20192,Sannomiya2019,Li2019,Roy2020,Tarun2020,Smitha2020,Sachdev1993,Maldacena2016,Rahmani20192,Yamada2020AKSL}.
However, physics arising from Majorana interactions in real Kitaev materials is still elusive.
On the other hand, various numerical studies on extended Kitaev model utilizing the spin representation, which effectively include
non-perturbative effects of Majorana interactions,
have been carried out~\cite{Chaloupka2010, Okamoto2013, Rau2014, Chaloupka2016, Nasu2017-2, Gohlke2018, Rusna2019,  Gordon2019, Jiang2019, Kaib2019, Lee2020, Gohlke2020, liu2020, buessen2021},
and nematicity is discussed in recent studies~\cite{Nasu2017-2, Gohlke2018, Lee2020, Gohlke2020, liu2020}.
Rich phase diagrams of Kitaev magnets are obtained from these studies based on the spin Hamiltonian,
while a further theoretical study is needed to understand the results of the recent specific heat measurements mentioned above,
which we believe is from the gapped QSL phase to another QSL phase~\cite{Tanaka2020}.

To focus on the nature of a QSL phase in Kitaev magnets from a different point of view,
we assume the flux-free state in our model.
Although the flux-free assumption
may be violated by 
strong additional spin-exchanges like Heisenberg, $\Gamma$, and $\Gamma^{\prime}$ terms,
it allows us not only to clarify properties of a QSL phase directly but also to explore topological phase transition mentioned above.
Besides, under the flux-free assumption we can combine the mean-field analysis and exact diagonalization,
which is impossible in the previous framework of the Majorana mean-field theory (see Refs.~\cite{Gohlke2018, Knolle2018}, for example),
so that we can study rotational symmetry breaking in the extended Kitaev model
from various aspects by computing the Majorana band structure,
the Majorana bond order, and the Chern number, which are hard to be handled by the direct treatment of the spin model.
Furthermore, the scenario that the Majorana four-body interactions give rise to rotational symmetry breaking is applicable to 
any flux sectors, which is described by interacting Majorana systems,
and thus, it is a good starting point to focus on one flux sector.
From now on, we elucidate that the four-body interactions can induce nematic-type Majorana bond ordered phase
which is a gapped QSL with zero Chern number, \textit{i.e.} the toric code phase.

\section{Model}
\begin{figure}[hbtp]
  \begin{center}
   \includegraphics[width=\columnwidth]{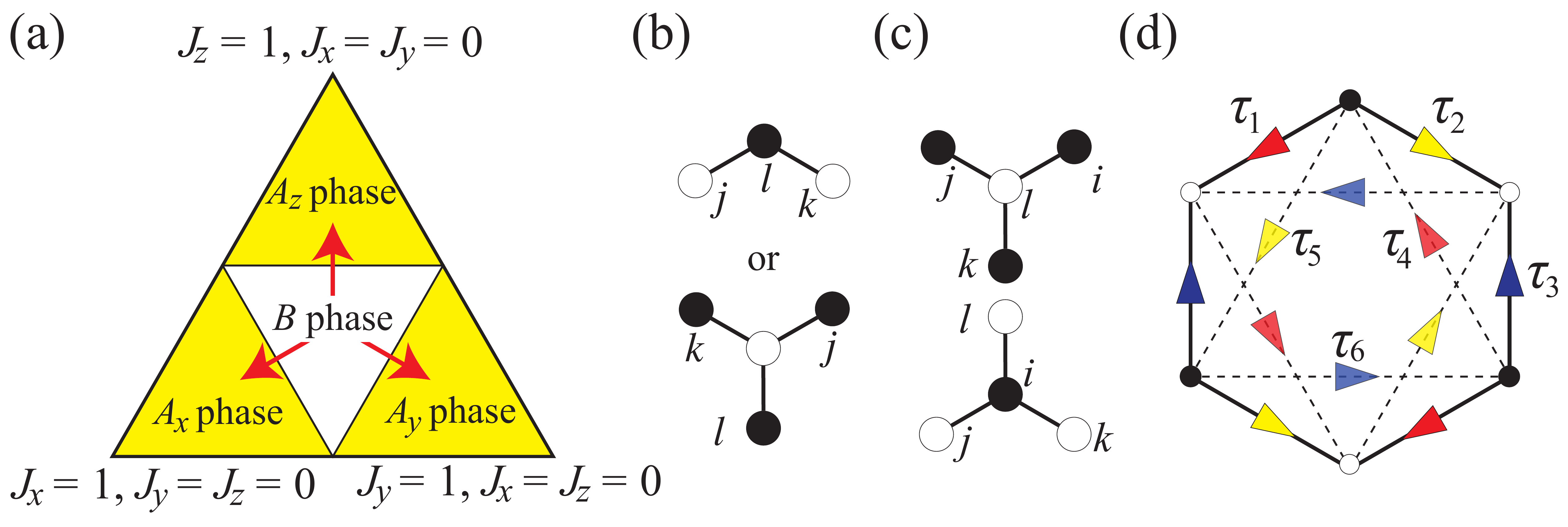}
\end{center}
    \caption{(a) Phase diagram of the Kitaev model. Red arrows indicate topological nematic phase transitions.
    (b) Configurations of three spins in Eq.~\eqref{Heff}.
  (c) Configurations of four Majorana fermions in Eq.~\eqref{Beff} and Eq.~\eqref{AAA}.
  (d) The definition of the hopping amplitude $\tau_a$. The arrows indicate the signs of the hopping terms.}
  \label{ph}
\end{figure}

The Hamiltonian of the pure Kitaev model is expressed as,
\begin{align}
    H=-J_x \sum_{\langle ij \rangle_x}\sigma_i^x\sigma_j^x-J_y \sum_{\langle ij \rangle_y}\sigma_i^y\sigma_j^y-J_z \sum_{\langle ij \rangle_z}\sigma_i^z\sigma_j^z,
\end{align}
where $\sigma_i$ represents the spin operator on the $i$th site
and $\langle ij \rangle$ means a nearest-neighbor (NN) bond.
$x$, $y$, and $z$ are indices for the bond direction.
The phase diagram in the parameter space $(J_x, J_y, J_z)$ obtained
in Kitaev's original paper is shown in Fig.~\ref{ph}(a).
For strongly anisotropic Kitaev interactions,
the so-called $A$ phase, which is the toric code phase, is realized.
On the other hand, for almost isotropic Kitaev interactions,
another phase ($B$ phase), which possesses gapless Dirac bands of Majorana fermions, appears.
Once the time-reversal symmetry is broken in $B$ phase, however, the spectrum acquires
an energy gap, leading to a chiral QSL state with the Chern number $\pm 1$.
For both phases, the system is described by Majorana fermions interacting with $Z_2$ gauge fields.
In the following, we restrict our analysis within the vortex-free states.
Then, in the case with applied magnetic fields, from the perturbation theory in the third order, 
effective interactions of three-spin operators are obtained:
\begin{align}
H_\textrm{eff}^{(3)}\sim-\frac{h_xh_yh_z}{J^2}\sum_{j,k,l}\sigma_j^x\sigma_k^y\sigma_l^z,
\label{Heff}
\end{align}
where $\bm{h}=(h_x,h_y,h_z)$ is an external magnetic field and
we here assume that $J_x=J_y=J_z=J$.
The site summation $i,j,k$ is taken as shown in Fig.~\ref{ph}(b).
In the Majorana representation, operators are rearranged to three types of terms:
\begin{align}
\sum_{j,k,l}\sigma_j^x\sigma_k^y\sigma_l^z\rightarrow-i\sum_{\langle\!\langle ij \rangle\!\rangle}c_ic_j+\sum_{\mathrm{Y}}c_ic_jc_kc_l+\sum_{\mathrm{Y}^{\prime}}c_ic_jc_kc_l,
\label{Beff}
\end{align}
where ``$\rightarrow$'' means taking the standard gauge in the 0-flux ground state, and $\langle\!\langle ij \rangle\!\rangle$ means a next-nearest-neighbor (NNN) bond.
The first term is an NNN hopping that gives rise to an energy gap, leading to the Chern number $\pm 1$, and
the others are four-Majorana interactions.
Here the site summation for $\sum_\mathrm{Y}$ is taken as shown in the upper Y-shaped diagram of Fig.~\ref{ph}(c)
and $\sum_\mathrm{Y^{\prime}}$ is taken for the lower one of Fig.~\ref{ph}(c).
In the following, we focus on these types of Majorana interactions for concreteness. 
However, we believe that our main results are generic also for other types of neighboring four-body interactions. 

It is noted that in the case with the off-diagonal exchange interaction,
$H_{\Gamma^{\prime}}=\Gamma^{\prime} \sum_{\langle ij\rangle_\alpha,\beta\neq\alpha} \left[\sigma_i^{\alpha}\sigma_j^{\beta}+\sigma_i^{\beta}\sigma_j^{\alpha}\right]
$ where $\alpha,\,\beta$ run for $x,$ $y,$ and $z$, the Y(Y')-shaped interactions also arise from the third-order perturbation terms of order $O(h_{x,y,z}{\Gamma^{\prime}}^2)$.
Besides, other non-Kitaev interactions, such as  Heisenberg and $\Gamma$ terms, contribute to bring additional NN and NNN hoppings as shown in Appendix~\ref{non-Kitaev}.
Thus, generally, the coefficients of the second and the third terms of Eq.~\eqref{Beff} are different from that of the first one, and
we end up with the following Majorana Hamiltonian,
\begin{align}
\mathcal{H}_\textrm{total} = \mathcal{H}_1+ \mathcal{H}_2+ \mathcal{H}_4,
\end{align}
with 
\begin{align}
\label{AAA}
\mathcal{H}_1 = it\sum_{\langle ij \rangle}c_ic_j,\quad \mathcal{H}_2 = i\kappa\sum_{\langle\!\langle ij \rangle\!\rangle}c_ic_j,\nonumber\\
\mathcal{H}_4 = g\left(\sum_{\mathrm{Y}}c_ic_jc_kc_l+\sum_{\mathrm{Y}^{\prime}}c_ic_jc_kc_l\right). 
\end{align}
We, here, stress again that the parameters $\kappa$ and $g$ are independent for real candidate materials of Kitaev magnets, as mentioned above.
The Majorana operator is represented as $c_i$ for the $i$th site, and the operators
obey $c_i^{\dagger} = c_i$ and $\{c_i,c_j\} = 2\delta_{ij}$. The hopping amplitude $t=J$ is
set unity without loss of generality.

\section{Mean-field analysis}
First, we apply the mean-field (MF) analysis to the four-body interaction term $\mathcal{H}_4$.
Then, the general form of the MF Hamiltonian is,
\begin{eqnarray}
\mathcal{H}_\mathrm{MF} = \sum_{a=1,2,3}i\tau_a\sum_{\langle ij \rangle_a}\eta_{ij}c_i c_j + \sum_{a=4,5,6}i\tau_a\sum_{\langle\!\langle ij \rangle\!\rangle_a}\eta_{ij}c_i c_j,
\label{Hmf}
\end{eqnarray}
where the index $a=1,\dots,6$ specifies the directions of hopping with the amplitude $\tau_a$, as
described in Fig.~\ref{ph}(d). The phase factors $\eta_{ij} = \pm1$ is necessary to
make $\mathcal{H}_\mathrm{MF}$ antisymmetric.
The sign of $\eta_{ij}$ is defined as illustrated in Fig.~\ref{ph}(d)
such that the direction of the arrow indicates the positive hopping from the $j$th site to the $i$th site.

Utilizing the Hellmann-Feynman theorem, we can derive self-consistent equations
as follows (see Appendix~\ref{MFA} for more details):
\begin{align}
\label{dada}
\tau_1 = \frac{t}{2}+g\Delta_4, \quad\tau_2 =\frac{t}{2}+g\Delta_5, \quad\tau_3 = \frac{t}{2}+g\Delta_6,\nonumber\\
\tau_4 = \frac{\kappa}{2}+\frac{g}{2}\Delta_1, \quad\tau_5 = \frac{\kappa}{2}+\frac{g}{2}\Delta_2, \quad\tau_6 = \frac{\kappa}{2}+\frac{g}{2}\Delta_3,
\end{align}
where the definition of bond order parameters is
\begin{align}
\Delta_a\equiv\braket{\Psi_\textrm{MF}|ic_i c_j|\Psi_\textrm{MF}}\quad(a=1,\dots,6),
\end{align}
with $\ket{\Psi_\mathrm{MF}}$ being the ground state of the MF Hamiltonian.

We solve the MF Hamiltonian by numerical iteration.
First, we focus on a nematic transition. 
The bond order parameters $\Delta_a$ ($a=1,\dots,6$) play an important role for nematic transitions with rotational symmetry breaking. 
The Majorana nematic phase considered here is defined as a phase with a bond order
in a specific direction. 
We define nematic order parameters as~\cite{Pujari2015},
 \begin{align}
 \phi &\equiv\Delta_1 + e^{2\pi i/3}\Delta_2 + e^{4\pi i/3}\Delta_3,
  \end{align}
   \begin{align}
 \psi &\equiv\Delta_4 + e^{2\pi i/3}\Delta_5 + e^{4\pi i/3}\Delta_6.
  \end{align}
If $\Delta_1=\Delta_2=\Delta_3$ ($\Delta_4=\Delta_5=\Delta_6$), $\phi$ ($\psi$) must be 0 from the definition,
but otherwise will have a nonzero value.
In other words, $\phi$ and $\psi$ characterizes breaking of the threefold rotational symmetry.

\begin{figure}[hbtp]
    \centering
    \includegraphics[width=\columnwidth]{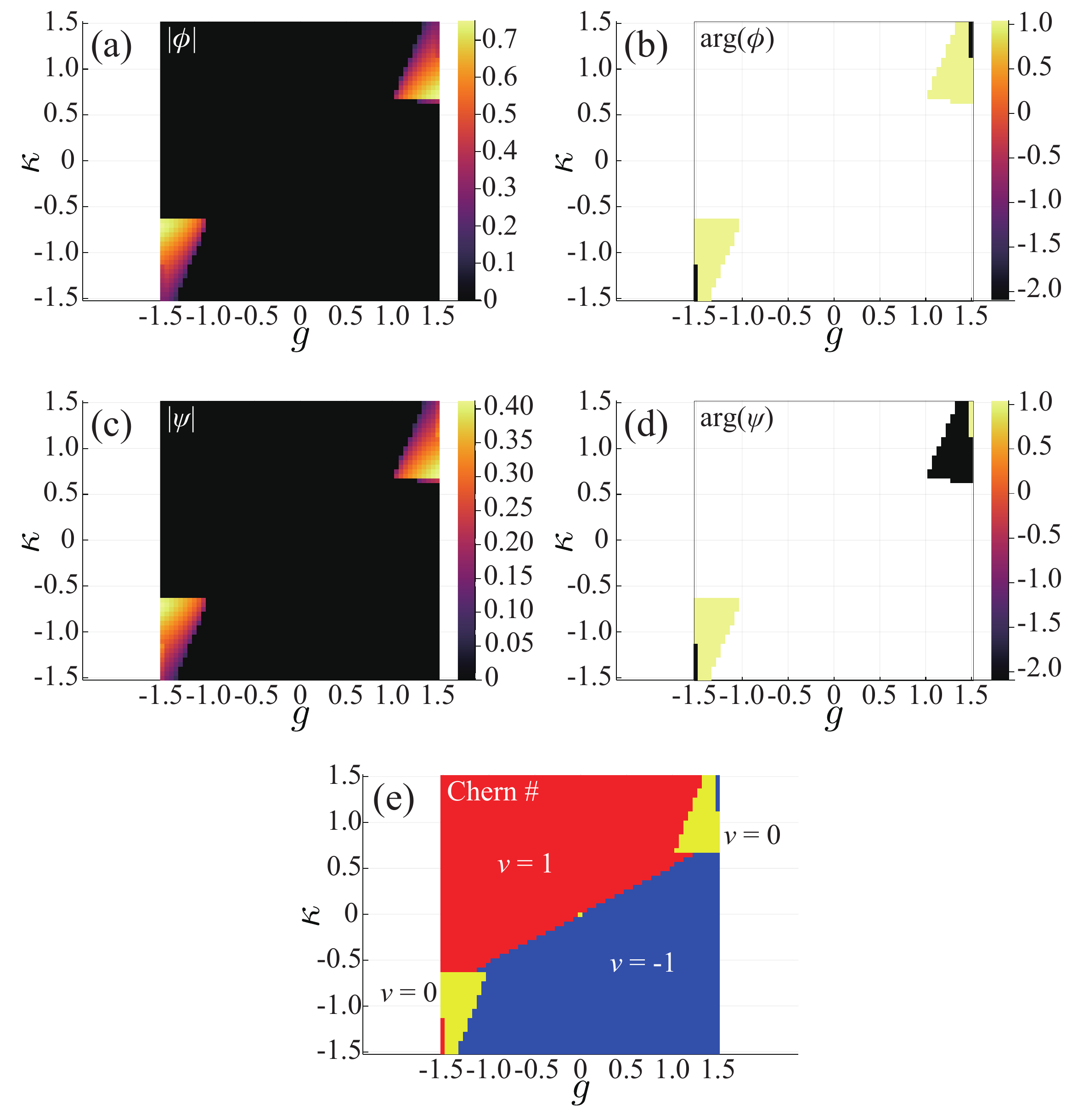}
  \caption{
  (a) Plot of $|\phi|$.
  (b) Plot of $\arg(\phi)$ in the regions with $|\phi|\neq 0$. $\arg(\phi)$ is $\pi/3$ in most of the regions,
  but becomes $-2\pi/3$ in the small areas around $|g|\sim 1.5$. 
 (c) Plot of $|\psi|$.
 (d) Plot of $\arg(\psi)$ in the regions with $|\psi|\neq 0$. 
  (e) The Chern number $\nu$ for each phase.
  }
  \label{fig2}
\end{figure}

As seen in Figs.~\ref{fig2}(a) and (c),  for large value of $|g|$, a nematic phase with $|\phi|\neq 0$, $|\psi|\neq 0$, characterized by two-fold rotational symmetry, appears.
We also find in Fig.~\ref{fig2}(b) that $\arg(\phi)$ is $\pi/3$ in most regions of the  $|\phi|\neq0$ area.
The argument of $\phi$ has informations about the direction of the nematic order.
Since $\Delta_a$ takes a negative value in this calculation, $\arg(\phi)=\pi/3$ corresponds to
the nematic order with $|\Delta_3| > |\Delta_1|, |\Delta_2|$.
On the other hand, as seen in Fig.~\ref{fig2}(d), $\arg(\psi)=\pi/3$ for $g<0$ in most of the nematic region, 
while  $\arg(\psi)=-2\pi/3$ for $g>0$. This is because that $\psi\rightarrow -\psi$ under time-reversal operation, while
$\phi \rightarrow \phi$.
In the calculations, we intentionally introduce tiny anisotropy of bonds to obtain symmetry-breaking solutions,
lifting three-fold degeneracy. 

Now, we examine topological features of the nematic phase.
For this purpose, 
we calculate the Chern number numerically using the Fukui-Hatsugai-Suzuki method~\cite{Fukui2005}. 
The results are shown in Fig.~\ref{fig2}(e).
In the chiral QSL phase without the nematic order, the Chern number $\nu=\pm1$.
Remarkably, on the other hand, in the nematic phase with $|\phi|\neq0$ and $|\psi|\neq 0$, we find $\nu=0$.
Thus, the nematic phase transition accompanies the topological phase transition.
This implies that the nematic transition which induces strong anisotropy of the Majorana hopping terms
($\tau_1$, $\tau_2$, $\tau_3$ in Eq.~\eqref{Hmf})
drives the system from $B$ phase to $A$ phase in the phase diagram shown in Fig.1(a).
In fact, in this nematic phase, since $Z_2$ vortices are still suppressed, the system has no magnetic long-range order, 
and is in the QSL state.
Moreover, the nematic phase is described by the quadratic Hamiltonian of Majorana fermions Eq.~\eqref{Hmf} with an energy gap. 
According to Kitaev's argument on the 16-way of anyons composed of $Z_2$ vortices and free fermions~\cite{Kitaev2006}, 
the gapful Majorana nematic phase with the Chern number $\nu=0$ is identified with the toric code phase with Abelian anyons.
It is noted that this nematic phase transition is the first-order type with the discontinuous changes 
of the order parameters $\phi$ and $\psi$, 
and hence, there is no gap closing at the transition point.
We stress that although the nonzero $\psi$ apparently causes the anisotropy of the Majorana hopping terms,
both of the two nematic orders $\phi$ and $\psi$ cooperatively stabilize 
the toric code phase, since $\phi$ and $\psi$ are nonlinearly coupled with each other  via the MF equations.
The details are shown in Appendix~\ref{N-ZN}.

At the end of this section, we should mention another phase realized in small regions with $\nu=\pm1$ inside nematic phases shown in Fig.2(e).
Although this phase is also a nematic phase with $|\phi|\neq 0$, $\arg(\phi)=-2\pi/3$ and hence,
$|\Delta_1|, |\Delta_2| > |\Delta_3|$ is realized. We call this phase ``zigzag nematic phase''.
This zigzag nematic phase is in $B$ phase, {\textit i.e.} with the Chern number $\nu=\pm 1$.

\section{Exact diagonalization}
To confirm the results of the MF calculation, 
we employ the exact diagonalization calculations. 
We apply the Lanczos method to the system size up to 32 sites. 
\begin{figure}[hbtp]
    \begin{center}
   \includegraphics[width=\columnwidth]{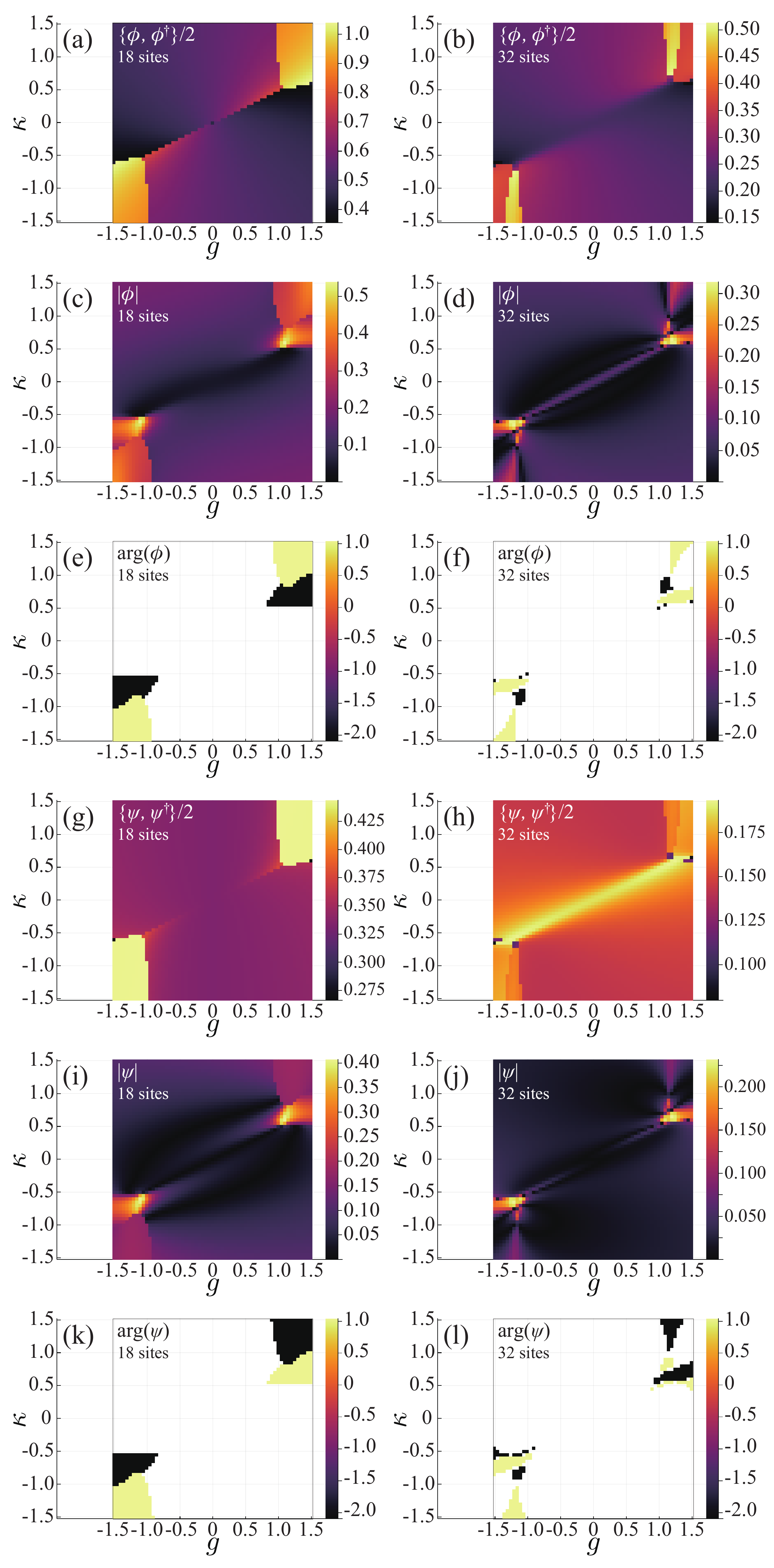}
    \end{center}
  \caption{The results obtained by exact diagonalization.
  (a)-(d) $\{\phi,\phi^\dagger\}/2$ (PBC),  and $|\phi|$ (APBC) for 18 sites and 32 sites.
  The orange and yellow regions in (c)  and (d) are identified with the nematic phases.
 (e)-(f)  $\arg(\phi)$ (APBC) for 18 sites and 32 sites.
  In yellow regions, $\arg(\phi)$ has $\pi/3$, 
 and in black regions, $\arg(\phi)=-2\pi/3$.
 (g)-(j) $\{\psi,\psi^\dagger\}/2$ (PBC) and $|\psi|$ (APBC) for 18 sites and 32 sites.
 (k)-(l) $\arg(\psi)$ (APBC) for 18 sites and 32 sites. Color convention is the same as (e), (f).
  }\label{fig3}
\end{figure}
First, we discuss the case of 18 sites.
When we use periodic boundary conditions (PBCs), $\phi$ and $\psi$ must be zero,
because the system completely preserves the rotational symmetry for finite size systems.
Thus, instead, we consider the fluctuation of the order parameters 
$\{\phi,\phi^\dagger\}/2$ and  $\{\psi,\psi^\dagger\}/2$ which can have a nonzero value even for the PBC.
The calculated results are shown in Figs.~\ref{fig3}(a) and (g). 
There are two bright yellow regions, which implies that
the nematic phase transition occurs discontinuously for large values of $g$.
This result confirms the MF calculation.
To examine the nematic order more directly, and to determine the
direction of bond orders, we switch to an antiperiodic boundary condition (APBC) to intentionally break the threefold
rotational symmetry so that we can compute the nonzero $|\phi|$ and $|\psi|$, and the arguments of $\phi$ and  $\psi$.
The nematic phase transition
is verified again as seen in Figs.~\ref{fig3}(c) and (i).
Moreover, 
another phase boundary appears within the nematic regions.
This implies that there are two different nematic phases.
One is a nematic phase which has one strong bond so that $\arg(\phi)=\pi+2n\pi/3\,(n\in \mathbb{Z})$,
and the other is a zigzag nematic phase with two strong bonds so that $\arg(\phi)=2n\pi/3\,(n\in \mathbb{Z})$ (see Fig.~\ref{fig3}(e)).
However, the shape of the phase boundary between two nematic phases is quite different from that of the MF result shown in Figs.~\ref{fig2}(b) and (d).
In fact, this phase boundary crucially depends on the system size, as seen below.

The results of the 32-site calculation are shown in Figs.~\ref{fig3}(b), (d), (f), (h), (j), and (l).
Although the region of the nematic phase depicted by orange color becomes narrower compared to the 18-site calculation,
the nematic phase is definitely realized for large values of $|g|$.
We also find two types of nematic phases as in the MF calculations and the 18-site calculations.
However, the shape of the phase boundary is quite different.  The region of the zigzag nematic phase with the Chern number $\nu=\pm 1$, where
$\arg(\phi)=-2\pi/3$, $\arg(\psi)=-2\pi/3$ ($\arg(\phi)=-2\pi/3$, $\arg(\psi)=\pi/3$) for $g<0$ ($g>0$),
becomes very small for the 32-site calculation.
To clarify the fate of the zigzag nematic phase, we need calculations for larger system sizes.
Although there is a significant finite-size effect in the calculations,
we can clearly observe the tendency towards the nematic phase transition.
In the end, small non-zero values in non-nematic phases seen in (c), (d), (i), and (j) are finite-size effects,
and decrease toward zero as the system size increases.

\section{Discussion}
Using the MF analysis and the exact diagonalization method, we find that
the topological nematic phase transition occurs in strongly correlated regions $|g| \sim t$, $g \sim \kappa$.
Since non-Kitaev interactions such as off-diagonal exchange interactions significantly renormalize $t$, $\kappa$, and $g$, 
it is possible to realize this condition for the extended Kitaev model.
Particularly, the $\Gamma$ term reduces the hopping amplitude $t$, and drives the system into strongly correlated regions.
We note that the Y(Y')-shaped interactions of order $\sim O(h_{x,y,z}{\Gamma^{\prime}}^2)$ cancel out for magnetic fields parallel to 
the honeycomb plane.
Nevertheless, other four-body interactions with different configurations of Majorana fermions, 
which do not disappear even for parallel magnetic fields,
can be generated from the Heisenberg exchange interaction and the other off-diagonal exchange interaction, {\it i.e.} the $\Gamma$ term.
The details of pertubative calculations with respect to non-Kitaev interactions are shown in Appendix~\ref{non-Kitaev} and
we expect that the nematic phase transition recently observed for $\alpha$-RuCl$_3$ under parallel magnetic fields~\cite{Tanaka2020} may be explained by taking into account these contributions.
To establish this scenario, we need further both theoretical and experimental investigations.
It is an interesting future issue to extend our analysis to general flux configurations, for which various Majorana phases are predicted~\cite{Zhang2019, Zhang2020}.
From an experimental side, an electric probe through the hyperfine interaction~\cite{yamada2020electric} may be useful for the detection of the Majorana nematic phase.

\section{Summary}
We have demonstrated that, in the Kitaev spin liquid state, four-body interactions among Majorana fermions which are induced by
applied magnetic fields and non-Kitaev exchange interactions give rise to the topological nematic phase transition from
the chiral QSL phase to the toric code phase.

\begin{acknowledgments}
We thank Y. Kasahara, Y. Matsuda, and T. Shibauchi for fruitful discussions.
This work was supported by JST CREST Grant No. JPMJCR19T5, Japan, and the Grant-in-Aid for Scientific Research on Innovative Areas
``Quantum Liquid Crystals (JP20H05163)'' from JSPS of Japan,
and JSPS KAKENHI (Grant No. ~JP17K05517, No.~JP20K03860 and No.~JP20H01857).
M.O.T. is supported by Program for Leading Graduate Schools: ``Interactive Materials Science Cadet Program''.
D.T. is supported by a JSPS Fellowship for Young Scientists and by JSPS KAKENHI Grant No. JP20J20385.
\end{acknowledgments}

\begin{appendix}
\section{MEAN-FIELD ANALYSIS AND SELF-CONSISTENT EQUATUIONS}
\label{MFA}
\begin{figure}[hbtp]
  \centering
  \includegraphics[width=40mm]{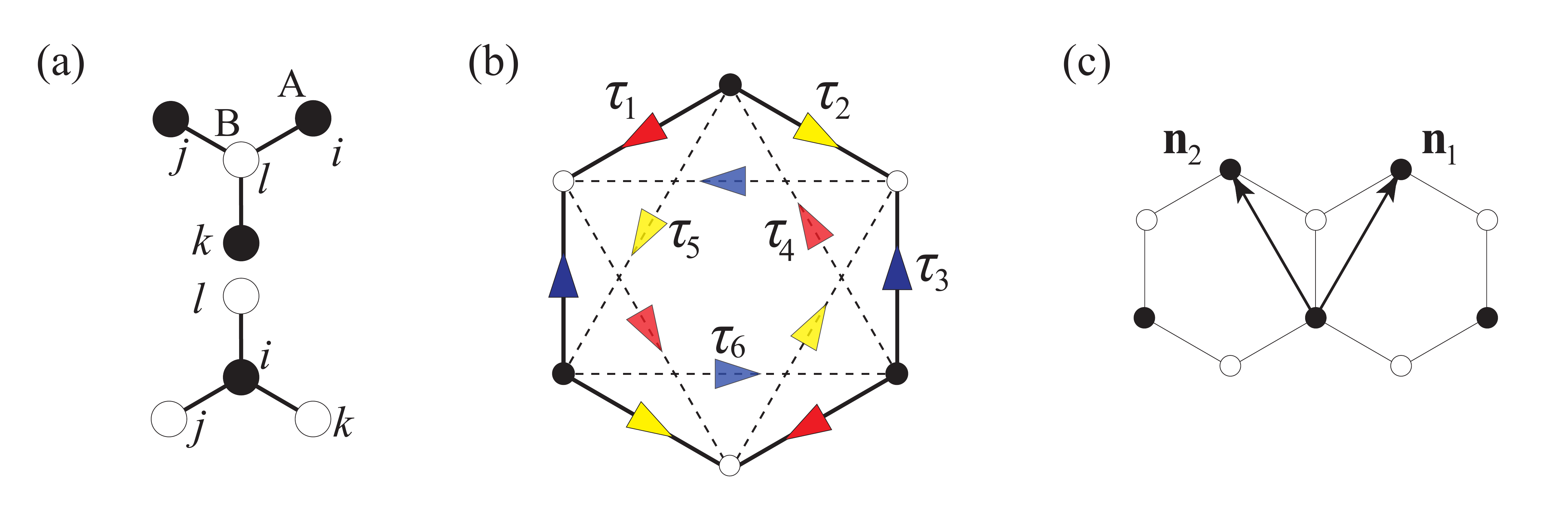}
    \caption{The fundamental translation vectors.}
  \label{SMfig1}
\end{figure}

In this section, we reveal the relation between the MF hopping strength $\tau_a$ and bond order $\Delta_a$.
The Fourier transform of a Majorana operator is given by,
\begin{align}
c_{\bm{k},\lambda}=\sqrt{\frac{1}{2N_{\textrm{unit}}}}\sum_{j}e^{-i\bm{r}_j\cdot\bm{k}} \,c_{j,\lambda}\quad(\lambda={\textrm{A, B}})
\end{align}
where $N_{\textrm{unit}}$ is the total number of unit cells and $j$ represents an each site. $\lambda$ is a position type inside the unit cell.
In other words, inverse Fourier transform is expressed as
\begin{align}
c_{j,\lambda}=\sqrt{\frac{2}{N_{\textrm{unit}}}}\sum_{\bm{k}}e^{i\bm{r}_j\cdot\bm{k}}\,c_{\bm{k},\lambda}
\end{align}
to satisfy $c_{\bm{k},\lambda}^{\dagger} = c_{-\bm{k},\lambda}$ and $\{c_{\bm{k},\lambda}, \,c_{\bm{q},\mu}^{\dagger}\}=\delta_{\bm{k},\bm{q}}\delta_{\lambda,\nu}$.
Using $\delta(\bm{k})=\sum_{j}e^{i\bm{r}_j\cdot\bm{k}} / N_{\textrm{unit}}$, we can calculate
each term of the MF Hamiltonian as,
\begin{eqnarray}
i\tau_1\sum_{\langle ij \rangle_1}\eta_{ij}c_i c_j
&=&i\tau_1\sum_{s}c_{(\bm{r}_s-\bm{n}_1,\,{\textrm A})} \,c_{(\bm{r}_s,\,{\textrm B)}}\,\,+\,\,{\textrm{H.c.}}\nonumber\\
&=&i\tau_1\sum_{s}\left[\sqrt{\frac{2}{N_{\textrm{unit}}}}\sum_{\bm{q}}e^{i(\bm{r}_s-\bm{n}_1)\cdot\bm{q}}\,c_{\bm{q},{\textrm A}}\right]\nonumber\\
&&\times\left[\sqrt{\frac{2}{N_{\textrm{unit}}}}\sum_{\bm{k}}e^{i\bm{r}_s\cdot\bm{k}}\,c_{\bm{k},{\textrm B}}\right]\,\,+\,\,{\textrm{H.c.}}\nonumber\\
&=&\sum_{\bm{k}}e^{i\bm{n}_1\cdot\bm{k}}c_{\bm{k},{\textrm A}}^{\dagger}c_{\bm{k},{\textrm B}}\,\,+\,\,{\textrm{H.c.}},
\end{eqnarray}
where we choose a basis $(\bm{n}_1,\bm{n}_2)$ of the translation group used in Ref.~\cite{Kitaev2006} (see Fig.~\ref{SMfig1}).
So the MF Hamiltonian in momentum space is obtained:
\begin{align}
\mathcal{H}_{\textrm{MF}} = \sum_{\bm{k}}\left(
    	\begin{array}{cc}
      		c_{\bm{k},{\textrm A}}^{\dagger} & c_{\bm{k},{\textrm B}}^{\dagger}
    	\end{array}
  	\right)
	\left(
    \begin{array}{cc}
      4D_2(\bm{k}) & 2D_1(\bm{k}) \\
    2 {D_1}^{\ast}(\bm{k}) & -4D_2(\bm{k})
    \end{array}
  \right)
  \left(
    	\begin{array}{c}
      		c_{\bm{k},{\textrm A}}\\
		c_{\bm{k},{\textrm B}}
    	\end{array}
  	\right),
\end{align}
where
\begin{align}
D_1(\bm{k}) = i\left(\tau_1 e^{i\bm{n}_1\cdot\bm{k}} +\tau_2 e^{i\bm{n}_2\cdot\bm{k}} + \tau_3\right),
\end{align}
\begin{eqnarray}
D_2(\bm{k})
&=& \tau_4\sin(-\bm{n}_2\cdot\bm{k})+\tau_5\sin(\bm{n}_1\cdot\bm{k})\nonumber\\
&&\qquad\qquad\qquad+\tau_6\sin((\bm{n}_2-\bm{n}_1)\cdot\bm{k}).
\end{eqnarray}
Then the energy spectrum and the occupied energy under the Fermi level are written as
\begin{align}
E(\bm{k}) = \pm 2\sqrt{|D_1(\bm{k})|^2+4D_2(\bm{k})^2},
\end{align}
\begin{align}
\label{eq1}
E_{\textrm{MF}}=-2\sum_{\bm{k}}\sqrt{|D_1(\bm{k})|^2+4D_2(\bm{k})^2},
\end{align}
where the summation $\sum_{\bm{k}}$ is taken over the first Brillouin zone.

Then, we evaluate the mean field energy by using the MF ground state $\ket{\Psi_{\textrm{MF}}}$ and the MF Hamiltonian $\mathcal{H}_{\textrm{MF}}$,
\begin{eqnarray}
E_{\textrm{MF}}&=&\langle\Psi_{\textrm{MF}}|\mathcal{H}_{\textrm{MF}}|\Psi_{\textrm{MF}}\rangle\nonumber\\
&=&\sum_{a=1,2,3}\left(\tau_a\sum_{\langle ij \rangle_a}\eta_{ij}\langle\Psi_{\textrm{MF}}|ic_i c_j|\Psi_{\textrm{MF}}\rangle\right)\nonumber\\
&&+\sum_{a=4,5,6}\left(\tau_a\sum_{\langle\!\langle ij \rangle\!\rangle_a}\eta_{ij}\langle\Psi_{\textrm{MF}}|ic_i c_j|\Psi_{\textrm{MF}}\rangle\right)\nonumber\\
&=&N\sum_{a=1,2,3}\tau_a\Delta_a+2N\sum_{a=4,5,6}\tau_a\Delta_a
\end{eqnarray}
where $N$ is the total number of sites, and we use,
\begin{eqnarray}
&\tau_1&\sum_{\langle ij\rangle_1}\eta_{ij}\langle\Psi_{\textrm{MF}}|ic_i c_j|\Psi_{\textrm{MF}}\rangle= N\tau_1\Delta_1,\nonumber
\end{eqnarray}
\begin{eqnarray}
&\tau_4&\sum_{\langle\!\langle ij\rangle\!\rangle_4}\eta_{ij}\langle\Psi_{\textrm{MF}}|ic_i c_j|\Psi_{\textrm{MF}}\rangle\nonumber\\
&&=\tau_4\sum_{\langle\!\langle ij\rangle\!\rangle_4}\eta_{ij}\langle\Psi_{\textrm{MF}}|ic_{i,{\textrm A}} c_{j,{\textrm A}}|\Psi_{\textrm{MF}}\rangle\nonumber\\
&&\qquad+\tau_4\sum_{\langle\!\langle ij\rangle\!\rangle_4}\eta_{ij}\langle\Psi_{\textrm{MF}}|ic_{i,{\textrm B}} c_{j,{\textrm B}}|\Psi_{\textrm{MF}}\rangle\nonumber\\
&&=2N\tau_4\Delta_4,\nonumber
\end{eqnarray}
respectively.
Utilizing the Hellmann-Feynman theorem, we obtain,
\begin{align}
\label{dd}
\Delta_a = \frac{1}{N}\frac{\partial E_{\textrm{MF}}}{\partial \tau_a}\,(a=1,2,3),\,\,\Delta_a = \frac{1}{2N}\frac{\partial E_{\textrm{MF}}}{\partial \tau_a}\,(a=4,5,6).
\end{align}
We can explicitly perform the derivatives using Eq.~\eqref{eq1} and get a set of equations,
\begin{align}
\Delta_1=-\frac{2}{N}\sum_{\bm{k}}\frac{\tau_1+\tau_2\cos((\bm{n}_2-\bm{n}_1)\cdot\bm{k})+\tau_3\cos(\bm{n}_1\cdot\bm{k})}{\sqrt{|D_1(\bm{k})|^2+4D_2(\bm{k})^2}},\nonumber
\end{align}
\begin{align}
\Delta_2=-\frac{2}{N}\sum_{\bm{k}}\frac{\tau_1\cos((\bm{n}_2-\bm{n}_1)\cdot\bm{k})+\tau_2+\tau_3\cos(\bm{n}_2\cdot\bm{k})}{\sqrt{|D_1(\bm{k})|^2+4D_2(\bm{k})^2}},\nonumber
\end{align}
\begin{align}
\Delta_3=-\frac{2}{N}\sum_{\bm{k}}\frac{\tau_1\cos(\bm{n}_1\cdot\bm{k})+\tau_2\cos(\bm{n}_2\cdot\bm{k})+\tau_3}{\sqrt{|D_1(\bm{k})|^2+4D_2(\bm{k})^2}},\nonumber
\end{align}
\begin{align}
\Delta_4=-\frac{4}{N}\sum_{\bm{k}}\frac{\sin(-\bm{n}_2\cdot\bm{k})D_2(\bm{k})}{\sqrt{|D_1(\bm{k})|^2+4D_2(\bm{k})^2}},\nonumber
\end{align}
\begin{align}
\Delta_5=-\frac{4}{N}\sum_{\bm{k}}\frac{\sin(\bm{n}_1\cdot\bm{k})D_2(\bm{k})}{\sqrt{|D_1(\bm{k})|^2+4D_2(\bm{k})^2}},\nonumber
\end{align}
\begin{align}
\Delta_6=-\frac{4}{N}\sum_{\bm{k}}\frac{\sin((\bm{n}_2-\bm{n}_1)\cdot\bm{k})D_2(\bm{k})}{\sqrt{|D_1(\bm{k})|^2+4D_2(\bm{k})^2}}.\nonumber
\end{align}

On the other hand, the total ground state energy which we will take variations with respect to $\{\tau_a\}$ can be easily written as
\begin{align}
\label{eq2}
\langle\mathcal{H}_{\textrm{total}}\rangle\equiv\langle\Psi_{\textrm{MF}}|\mathcal{H}_1+\mathcal{H}_2+\mathcal{H}_4|\Psi_{\textrm{MF}}\rangle,
\end{align}
and one can express  the first term and the second term with $\Delta_a$ as
\begin{align}
\langle\mathcal{H}_1\rangle=N\times \frac{t}{2}(\Delta_1+\Delta_2+\Delta_3),\nonumber
\end{align}
\begin{align}
\langle\mathcal{H}_2\rangle=N\times \kappa(\Delta_4+\Delta_5+\Delta_6).\nonumber
\end{align}
In addition, applying Wick's theorem,
the final term of Eq.~\eqref{eq2} can be written as
\begin{align}
\sum_{\textrm Y}\langle c_i c_j c_k c_l\rangle =\frac{N}{2}\left(\Delta_1\Delta_4+\Delta_2\Delta_5+\Delta_3\Delta_6\right),\nonumber
\end{align}
\begin{align}
\sum_{\textrm Y'}\langle c_i c_j c_k c_l\rangle =\frac{N}{2}\left(\Delta_1\Delta_4+\Delta_2\Delta_5+\Delta_3\Delta_6\right).\nonumber
\end{align}
In short, the total energy density can be expressed as,
\begin{eqnarray}
\frac{\langle\mathcal{H}_{\textrm{total}}\rangle}{N}
&=&\frac{t}{2}(\Delta_1+\Delta_2+\Delta_3)+\kappa(\Delta_4+\Delta_5+\Delta_6)\nonumber\\
&&\qquad+g\left(\Delta_1\Delta_4+\Delta_2\Delta_5+\Delta_3\Delta_6\right).
\end{eqnarray}
Now, we can obtain the self-consistent equations by minimizing the energy with respect to variational parameters $\{\tau_a\}$,
\begin{align}
\frac{\partial \langle\mathcal{H}_{\textrm{total}}\rangle}{\partial \tau_a}=0,
\end{align}
or more explicitly,
\begin{align}
\label{ee}
(\frac{t}{2}+g\Delta_4)\frac{\partial \Delta_1}{\partial \tau_a}+(\frac{t}{2}+g\Delta_5)\frac{\partial \Delta_2}{\partial \tau_a}+(\frac{t}{2}+g\Delta_6)\frac{\partial \Delta_3}{\partial \tau_a}\qquad\quad\nonumber\\
+(\kappa+g\Delta_1)\frac{\partial \Delta_4}{\partial \tau_a}+(\kappa+g\Delta_2)\frac{\partial \Delta_5}{\partial \tau_a}+(\kappa+g\Delta_3)\frac{\partial \Delta_6}{\partial \tau_a}=0.
\end{align}
Finally, comparing Eq.~\eqref{dd} and Eq.~\eqref{ee}, we end up with,
\begin{align}
\tau_1 = \frac{t}{2}+g\Delta_4,\quad\tau_2 =\frac{t}{2}+g\Delta_5,\quad\tau_3 = \frac{t}{2}+g\Delta_6,\nonumber
\end{align}
\begin{align}
\tau_4 = \frac{\kappa}{2}+\frac{g}{2}\Delta_1,\quad\tau_5 = \frac{\kappa}{2}+\frac{g}{2}\Delta_2,\quad\tau_6 = \frac{\kappa}{2}+\frac{g}{2}\Delta_3.
\label{taua}
\end{align}

\section{NEMATIC PHASE AND ZIGZAG NEMATIC PHASE}
\label{N-ZN}
\begin{figure}[hbtp]
  \centering
  \includegraphics[width=\columnwidth]{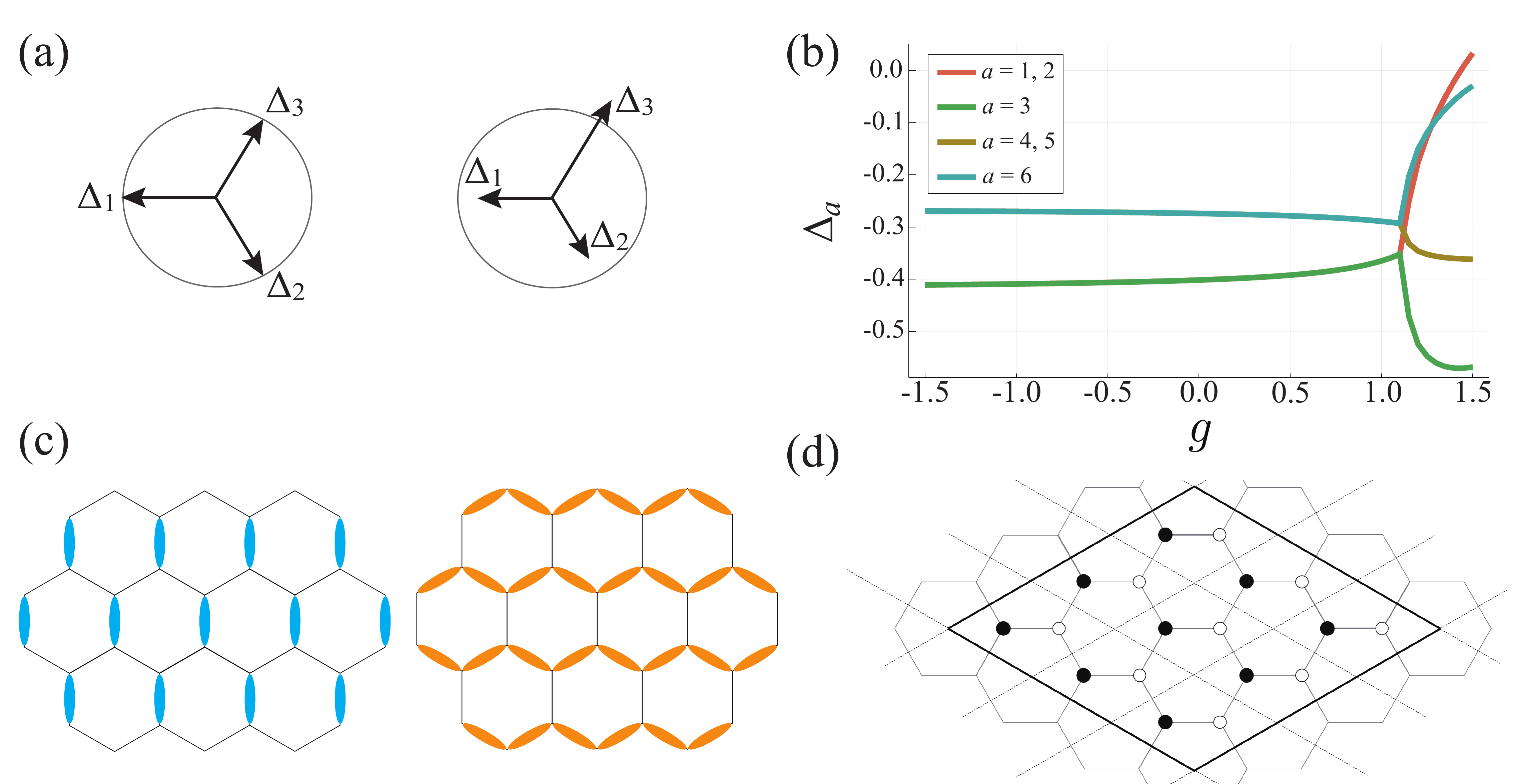}
    \caption{(a) A schematic view of the nematic order parameter $\phi$.
    The left figure corresponds to the case of $\Delta_1=\Delta_2=\Delta_3$ so that $\phi$ has $0$.
    On the other hand, $\phi$ has a nonzero value when the rotational symmetry is broken as shown in the right figure.
    (b) $\Delta_a$ versus $g$ for the case of $\kappa=1.0$.
    All of $\Delta_a$ change discontinuously at the transition point.
    (c) Schematic views of a nematic phase with one strong bond order (left) and a zigzag nematic phase which has strong bonds in two directions (right).
    (d) the clusters used in exact diagonalization with 18 sites. The threefold rotational symmetry is preserved when periodic boundary condition is applied.}
  \label{SMfig2}
\end{figure}

In this section, we discuss how to distinguish between a nematic phase and a zigzag nematic phase in strongly interacting regions.
Since $\Delta_a\,(a=1,...,6)$ has a negative value in most of parameter regions (see Fig.~\ref{SMfig2}(b)),
 the nematic order of $\phi$ is schematically expressed as shown in Fig.~\ref{SMfig2}(a); 
$\arg(\phi)$ reflects the direction of bond orders with $|\phi|\neq0$.
For instance, $\arg(\phi) =\pi/3$ implies $|\Delta_1|=|\Delta_2|<|\Delta_3|$, which is a direct signal of a nematic phase with one strong bond along the
$a=3$ direction.
On the other hand, if $\arg(\phi) =-2\pi/3$, there are two strong bonds in $a=1,2$ directions.
In general, $\arg(\phi) =\pi+2n\pi/3\,(n\in\mathbb{Z})$ represents a nematic phase in a one-strong-bond order
and $\arg(\phi) = 2n\pi/3\,(n\in\mathbb{Z})$ indicates a zigzag nematic phase that has two strong bonds (see Fig.~\ref{SMfig2}(c)).
Furthermore, Fig.~\ref{SMfig2}(b) shows that the symmetry breakings of $\phi$ and $\psi$ occur at the same transition point,
where we intendedly realize anisotropy of bond orders by substituting several patterns of initial values and take the lowest energy state in each calculation.

We also note symmetrical properties of $\phi$ and $\psi$ under time reversal operation.
Since time-reversal operation for Majorana operators depends on the sublattice degrees of freedom,
$\Delta_a$ for $a=1,2,3$ preserves the time-reversal symmetry, while $\Delta_a$ for $a=4,5,6$ changes its sign under time-reversal operation.
Therefore, the time-reversal operation acts on $\phi$ and $\psi$ as described below:
\begin{align}
\phi\rightarrow\phi'=\phi,\quad \psi\rightarrow\psi'=-\psi.
\end{align}
Because of these symmetrical properties, $\arg(\phi)$ is not changed by the transformation $g\rightarrow -g$ and $\kappa \rightarrow -\kappa$,
which correspond to the time-reversal operation, while $\arg(\psi)$ changes its value as $\pi/3 \leftrightarrow -2\pi/3$
 under this transformation, as seen in Figs.~2 and 3 in the main text.
In short, for $g<0$, the region with $\arg(\psi)=\pi/3$ ($\arg(\psi)=-2\pi/3$) corresponds to $\Delta_4=\Delta_5<\Delta_6$ ($\Delta_4=\Delta_5>\Delta_6$),
and for $g>0$, the region with $\arg(\psi)=-2\pi/3$ ($\arg(\psi)=\pi/3$) corresponds to $\Delta_4=\Delta_5<\Delta_6$ ($\Delta_4=\Delta_5>\Delta_6$).

\section{DERIVATION OF FOUR-MAJORANA INTERACTIONS ARISING FROM NON-KITAEV INTERACTIONS}
\label{non-Kitaev}
In this section, we present the details of the derivation of the four-Majorana interactions caused by non-Kitaev interactions on the basis of
perturbative calculations. 
In these calculations, we need to evaluate matrix elements of perturbation terms for the eigenstates of $Z_2$ vortices.
For this purpose, we, first, examine the operation of spin operators on the eigenstate, and then, 
explore for nonzero matrix elements of perturbation terms generated by symmetric off-diagonal exchange interactions, {\textit i.e.} the $\Gamma$ term, and the $\Gamma'$ term, and also by the Heisenberg exchange interaction. 
It is found that the Y-shaped four-Majorana interactions are generated by the $\Gamma'$ term combined with the Zeeman term.
On the other hand, the $\Gamma$ term and the Heisenberg interaction give rise to armchair-shaped interaction, and zigzag-shaped interactions,
respectively (see below).
Up to the third-order perturbation, this analysis exhausts all four-body interactions among neighboring Majorana fermions generated
by the combination of applied magnetic fields and the Heisenberg or the symmetric off-diagonal exchange interactions.
\subsection{Flip of $Z_{2}$ gauge fields due to the operation of spin operators} 
In the following, we use the Majorana representation of $s=1/2$ spin operators introduced by Kitaev~\cite{Kitaev2006},
$\sigma^x_j=ib^x_jc_j$, $\sigma^y_j=ib^y_jc_j$, $\sigma^z_j=ib^z_jc_j$. 
The pure Kitaev model is expressed in terms of itinerant Majorana fields $c_j$ interacting with $Z_2$ gauge fields $\hat{u}^{\alpha}_{jk}=ib^{\alpha}_jb^{\alpha}_k$ on $\alpha$-bonds ($\alpha=x,y,z$).
Here, we consider effects of the operation of spin operators on the eigenstate of flux operators, $\hat{w}$,
\begin{eqnarray}
\sigma^{z}_{j} \ket{\Psi_{w}} &=& \Pi_{k} \frac{1+D_{k}}{2} \ket{\phi},\\
\ket{\Psi_{w}} &\equiv& \Pi_{k} \frac{1+D_{k}}{2} \ket{\Psi_{u}},\\
\ket{\phi} &\equiv& {\sigma_{j}}^{z} \ket{\Psi_{u}}.
\end{eqnarray}
where $\ket{\Psi_{w}}$ and $\ket{\Psi_{u}}$ are, respectively, the eigenstates of flux operators, and that of $Z_{2}$ gauge fields $\hat{u}$,
and $D_k=b^x_kb^y_kb^z_kc_k$.
We also define a state $\ket{\phi}$ that is an excited state with a flipped $Z_{2}$ gauge field.
Note that the following relations hold:
\begin{eqnarray}
\comm{\sigma^{z}}{D}&=& \comm{ib^{z}c}{b^{x}b^{y}b^{z}c}=0,\\
\comm{{\sigma_i}^{\alpha}}{\hat{u}^{\beta}_{kl} }&=& \comm{i {b_i}^{\alpha}c_i}{i {b_k}^{\beta}{b_l}^{\beta}}\nn\\
&=&-2\delta_{\alpha\beta}\delta_{ik}{b_{l}}^{\alpha}c_{i}+2\delta_{\alpha\beta}\delta_{il}{b_{k}}^{\alpha}c_i.
\end{eqnarray}
Using these relations, one can see that $\ket{\phi}$ is actually a state with a flipped eigenvalue of $\hat{u}^z$:
\begin{eqnarray}
\uz_{jk}\ket{\phi} &=& \uz_{jk} {\sigma_{j}} ^{z}\p\nn\\
&=&   - {\sigma_{j}} ^{z}\uz_{jk}\p\nn\\
&=& (- u^{z}_{jk}) \ket{\phi}.
\end{eqnarray}
This means that the sign of the eigenvalue of the $Z_{2}$ gauge field for the $\ket{\phi}$ state is opposite to that of the $\p$ state. 
The flip of the $Z_{2}$ gauge field  occurs also  in the case that the spin operator $\sigma^z$ acts on the other site of the $z$-bond,
\begin{eqnarray}
|\phi^{\prime}\rangle &\equiv& {\sigma_{k}}^{z} \ket{\Psi_{u}},\\
\uz_{jk}|\phi^{\prime}\rangle &=& \uz_{jk} {\sigma_{k}} ^{z}\p\nn\\
&=&   - {\sigma_{k}} ^{z}\uz_{jk}\p\nn\\
&=& (- u^{z}_{jk}) |\phi^{\prime}\rangle.
\end{eqnarray}

\subsection{Perturbative calculations with respect to non-Kitaev interactions} 
We start with the following Hamiltonian for candidate materials of the Kitaev magnet on a honeycomb lattice such as $\alpha$-RuCl$_3$ and Na$_2$IrO$_3$ \cite{Rau2014,PhysRevLett.113.107201,Winter2016,PhysRevB.96.054410}, 
\begin{eqnarray}
\mathcal{H}=\mathcal{H}_{K}+\mathcal{H}_{J_{H}}+\mathcal{H}_{\Gamma}+\mathcal{H}_{\Gamma'},
\end{eqnarray}
\begin{eqnarray}
\mathcal{H}_{K}=-J\sum_{ \langle ij\rangle_{\alpha}} \sigma_i^{\alpha}\sigma_j^{\alpha},  \label{eq:hamk}
\end{eqnarray} 
\begin{eqnarray}
\mathcal{H}_{J_{H}}=J_{H}\sum_{\langle ij\rangle} \bm{\sigma}_i\cdot\bm{\sigma}_j, \label{eq:hamh}
\end{eqnarray}
\begin{eqnarray}
\mathcal{H}_{\Gamma}=\Gamma\sum_{\substack{ \langle ij\rangle_{\alpha} \\ \beta,\gamma \neq \alpha}} [\sigma_i^{\beta}\sigma_j^{\gamma}+\sigma_i^{\gamma}\sigma_j^{\beta}], \label{eq:hamgam1}
\end{eqnarray}
\begin{eqnarray}
\mathcal{H}_{\Gamma'}=\Gamma'\sum_{\substack{  \langle ij\rangle_{\alpha} \\ \beta \neq \alpha}} [\sigma_i^{\alpha}\sigma_j^{\beta}+\sigma_i^{\beta}\sigma_j^{\alpha}],  \label{eq:hamgam2}
\end{eqnarray}
where $\sigma^{\alpha}_i$ is an $\alpha=x,y,z$ component of an $s=1/2$ spin operator at a site $i$.
$\mathcal{H}_{K}$ is the Kitaev interaction, 
$\mathcal{H}_{J_{H}}$ is the Heisenberg exchange interaction between the nearest neighbor sites, and
$\mathcal{H}_{\Gamma}$ and $\mathcal{H}_{\Gamma'}$ are symmetric off-diagonal exchange interactions.
Here, $\langle ij\rangle_\alpha$ denotes that the $i$th site and the $j$th site are connected via a nearest-neighbor $\alpha$-bond on the honeycomb lattice.
We note that the definitions of $J_H$, $\Gamma$, and $\Gamma'$ are different from conventional ones used in first principle calculations 
\cite{Rau2014,Winter2016,PhysRevB.96.054410} by a factor $1/4$.
We also take into account the Zeeman term due to an external magnetic field $\bm{h}=(h_x,h_y,h_z)$,
\begin{eqnarray}
\mathcal{H}_Z=-\sum_i [h_x\sigma^x_i+h_y\sigma^y_i+h_z\sigma^x_i].
\end{eqnarray}

Putting $V^{\prime}=\mathcal{H}_{J_{H}}+\mathcal{H}_{\Gamma}+\mathcal{H}_{\Gamma'}+\mathcal{H}_{Z}$,
we carry out the perturbative expansions with respect to $V'$ around the vortex-free ground state in the same spirit as the Kitaev's paper~\cite{Kitaev2006}.
In this perturbation analysis, intermediate excited states have vortex excitations (visons) with finite energy gaps.
On the basis of the results derived in the previous section, we examine nonzero matrix elements in each perturbation term which generates 
four-Majoarana interactions.

\begin{figure}[http]
 \centering
   \includegraphics[width=\columnwidth]{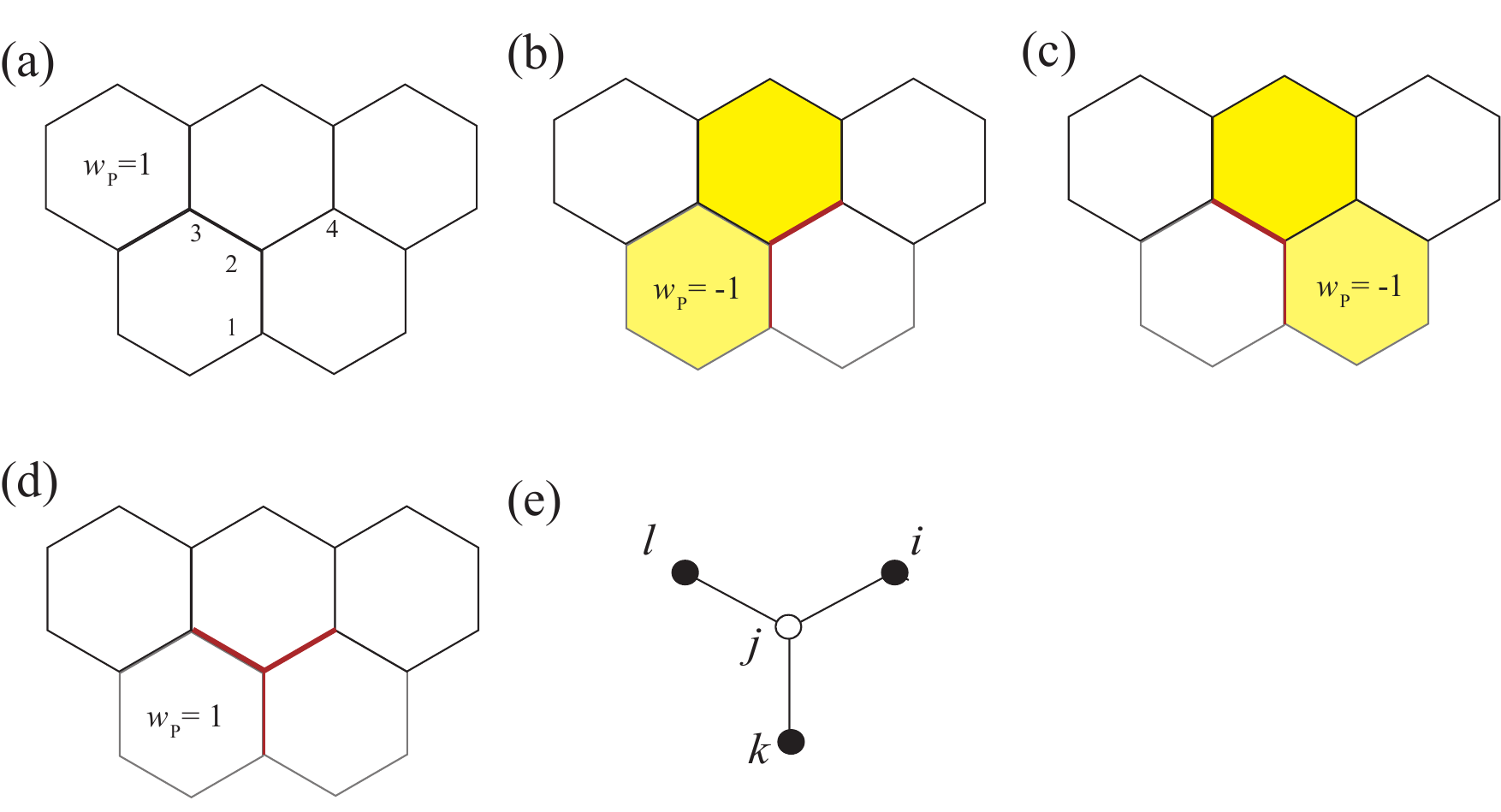}
 \caption{The operation of $\Gamma' \sigma^{x}_{2}\sigma^{z}_{1}$  on the vortex-free ground state
 shown in (a)
 flips
 the $Z_2$ gauge fields on the bonds denoted by red color shown in (b), and generates
 two visons (yellow hexagons). 
 The operation of $\Gamma' \sigma^{x}_{2}\sigma^{y}_{3}$ on the state shown in (b) 
 results in the excited state with two visons as shown in (c). 
 The operation of $h_x\sigma^{x}_{4}$ on the state shown in (c) results in the vortex-free ground state shown in (d). 
 (e) An example of the configuration of the Y-shaped interaction.} 
 \label{takahashi-supp1}
\end{figure}

We, first, consider the Y-shaped interaction, which is derived from the third-order perturbation with respect to the $\Gamma'$ term, $\mathcal{H}_{\Gamma'}$, and the Zeeman term $\mathcal{H}_Z$.
To be concrete, we show an example of the perturbation processes in Fig.~\ref{takahashi-supp1}.
In this example of the third-order perturbations, the operations of $\Gamma' \sigma^{z}_{1}\sigma^{x}_{2}$ and $\Gamma' \sigma^{x}_{2}\sigma^{y}_{3}$ and $h_x\sigma^{x}_{4}$
on the vortex-free ground state result in the final state, which is also the vortex-free ground state.
Thus, these perturbation processes are allowed within the ground state sector.
On the other hand, $\mathcal{H}_{\Gamma}$ and $\mathcal{H}_{J_{H}}$ do not give the perturbation processes within the ground state sector
which result in the Y-shaped interaction.
Thus, we, here, omit these two terms in $V'$.
Then, the third-order perturbation term which leads to the Y-shaped interaction is given by,
\begin{eqnarray}
\mathcal{H}^{(3)}_{\textrm Y}&=&\Pi_{0}V^{\prime}G'_{0}(E)V^{\prime}G'_{0}(E)V^{\prime}\Pi_{0},\nn\\
&=&-\frac{3{\Gamma^{\prime}}^{2}}{\Delta^2}
 {\displaystyle \sum_{\begin{subarray}{c}   \alpha = x,y,z  \\ \beta \neq \alpha \\ \gamma \neq \beta, \alpha
 \\ \end{subarray}}
 \sum_{\begin{subarray}{c}  \langle ij\rangle_{\alpha} \\ \langle jl\rangle_{\gamma} \\ \langle jk\rangle_{\beta} \\ \end{subarray}}}   
 \Pi_0\sigma^{\alpha}_{i}\sigma^{\gamma}_{l}\sigma^{\beta}_{k}(h_\beta+h_\gamma)\Pi_0,\label{eq:yshaped}
\end{eqnarray}
where $\alpha,\beta,\gamma = x,y,z$, and $\Pi_{0}$ is a projection to the vortex-free spin liquid state.
Also, $\Delta$ is the energy gap of two visons, which is given by $\Delta\sim0.26J$ \cite{Kitaev2006} for the configurations shown in Figs.~\ref{takahashi-supp1}(b), (c).
To analyze this term more precisely, we use the fact that in the Majorana fermion representation of the Kitaev spin liquid state, 
gauge Majorana fields $b_i^{\alpha}$ should be paired on the $\alpha$-bond connecting two sites $i$ and $j$ to form $Z_2$ gauge fields $\hat{u}_{ij}^{\alpha}={\textrm i}b^{\alpha}_ib^{\alpha}_j$, since the Kitaev spin liquid state is expressed by the eigen state of the $Z_2$ gauge fields.
Then in the case of $\alpha = x$, Eq.~\eqref{eq:yshaped} is recast into,
\begin{eqnarray}
\mathcal{H}^{(3)}_{\textrm Y}&=\Pi_0\Biggl[
{\displaystyle \sum_{\begin{subarray}{c} \langle ij\rangle_{x} \\ \langle jl\rangle_{y}  \\ \langle jk\rangle_{z} 
 \\ \end{subarray}}}
 (-\frac{6{\Gamma^{\prime}}^{2}(h_y+h_z)}{\Delta^2})c_{i}c_{l}c_{k}c_{j}\hat{u}^{x}_{ij}\hat{u}^{y}_{lj}\hat{u}^{z}_{kj}\nn
\Biggr]\Pi_0. \label{eq:yshpaedex}
\end{eqnarray}
This amounts to the Y-shaped interaction of four itinerant Majorana fermions on the sites $i$, $j$, $k$, $l$ shown in Fig.~\ref{takahashi-supp1}(e).

\begin{figure}[http]
 \centering
   \includegraphics[width=\columnwidth]{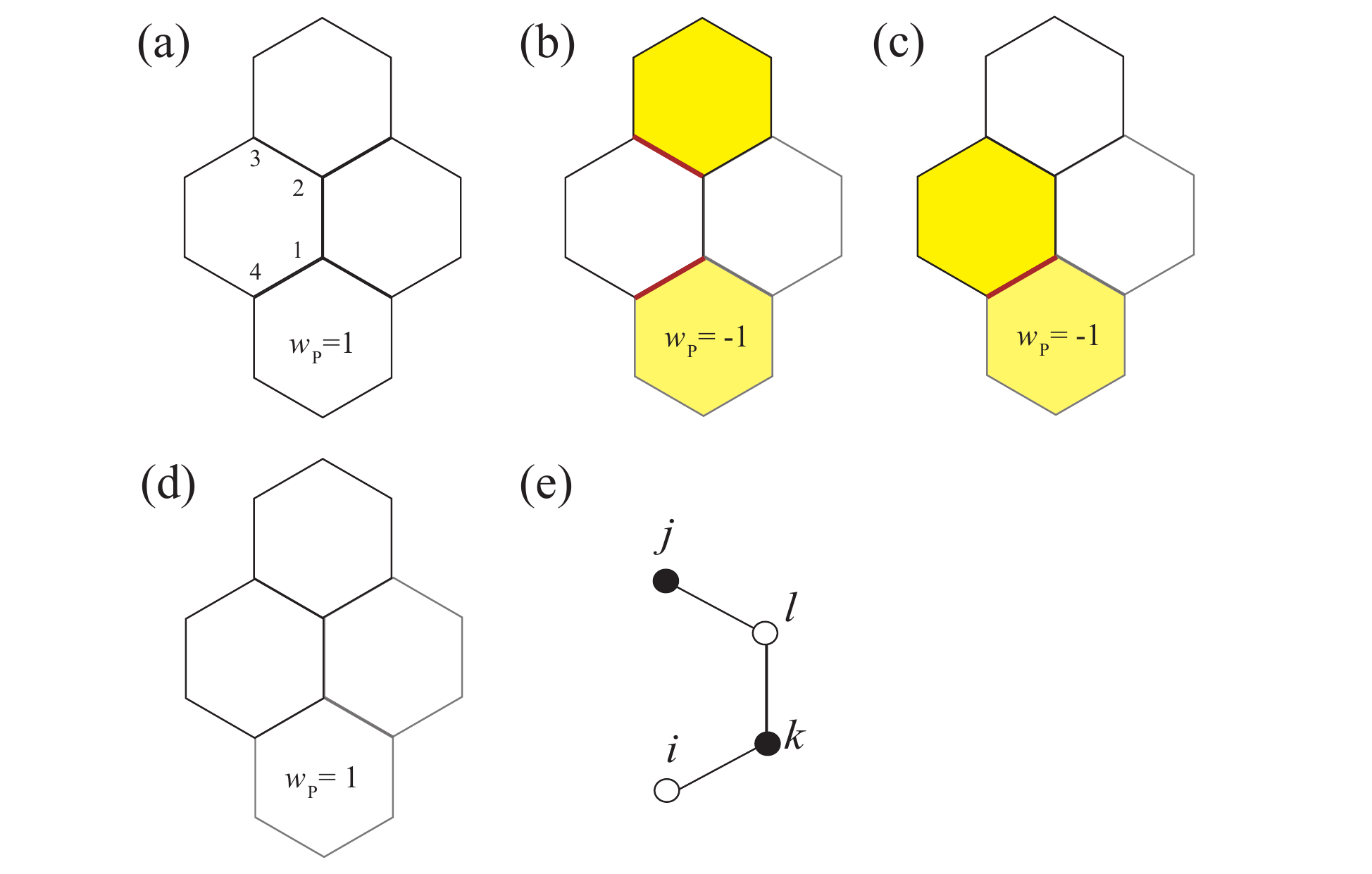}
 \caption{The operation of $\Gamma \sigma^{x}_{1}\sigma^{y}_{2}$  on the vortex-free ground state
 shown in (a)
 flips
 the $Z_2$ gauge fields on the bonds denoted by red color shown in (b), and generates
 two visons. 
 The operation of $h_y\sigma^{y}_{3}$ on the state shown in (b) 
 results in the excited state with two visons as shown in (c). 
 The operation of $h_x\sigma^{x}_{4}$ on the state shown in (c) results in the vortex-free ground state shown in (d). 
 (e) An example of the configuration of the armchair-shaped interaction.}
 \label{takahashi-supp2}
\end{figure}

\begin{figure}[http]
 \centering
 \includegraphics[width=\columnwidth]{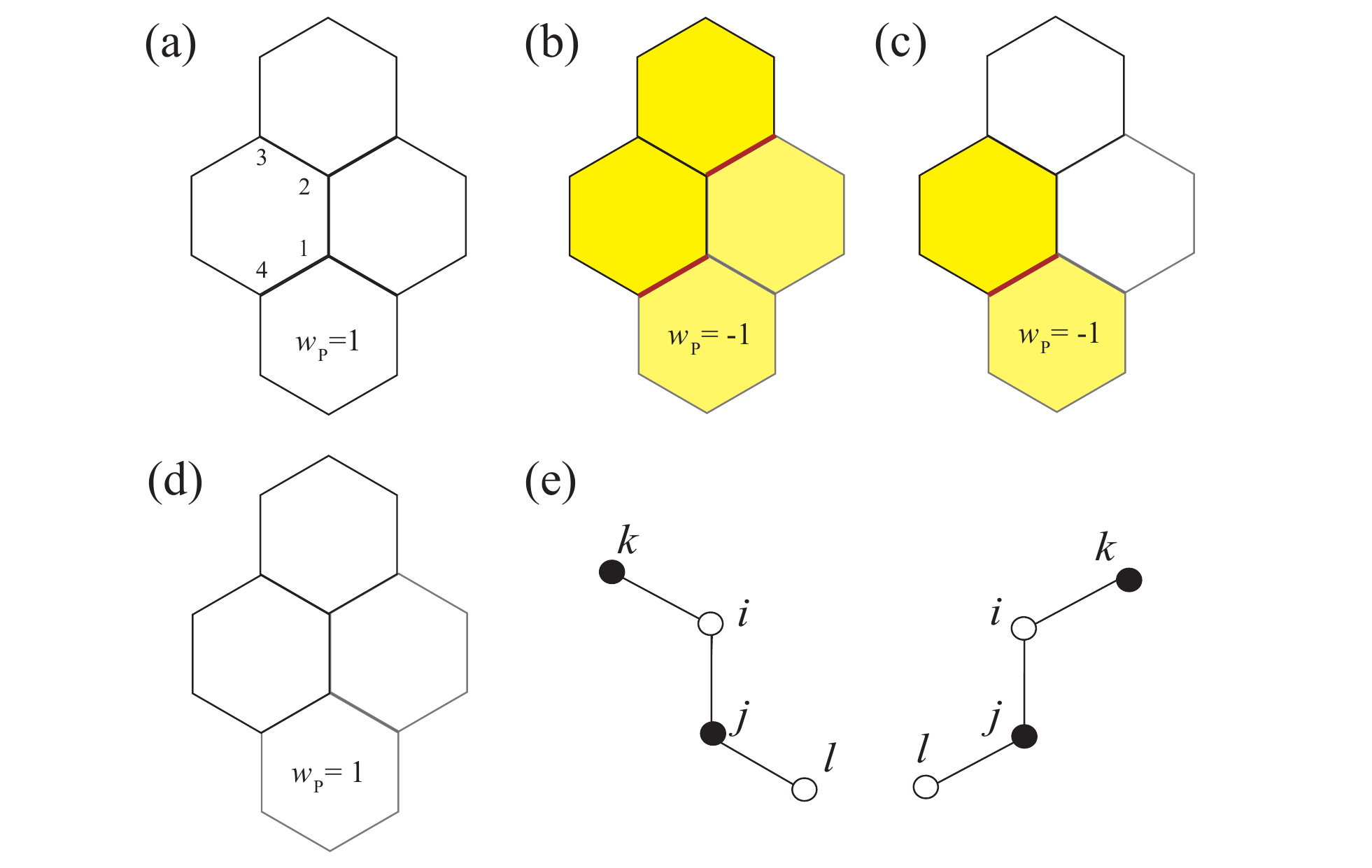}
 \caption{The operation of $J_H \sigma^{x}_{1}\sigma^{x}_{2}$  on the vortex-free ground state
  shown in (a)
 flips
 the $Z_2$ gauge fields on the bonds denoted by red color shown in (b), and generates
 four visons. 
 The operation of $h_x\sigma^{x}_{3}$ on the state shown in (b) 
 results in the excited state with two visons as shown in (c). 
 The operation of $h_x\sigma^{x}_{4}$ on the state shown in (c) results in the vortex-free ground state shown in (d). 
 (e) Examples of the configuration of the zigzag-shaped  interaction.}
 \label{takahashi-supp3}
\end{figure}

Secondly, we consider the armchair-shaped interaction, which is generated by the third-order perturbation with respect to the $\Gamma$ term 
and the Zeeman term.
An example of the perturbation processes is shown in Fig.~\ref{takahashi-supp2}.
In this example of the third-order perturbations, the operations of $\Gamma \sigma^{x}_{1}\sigma^{y}_{2}$ and $h_y\sigma^{y}_{3}$ and $h_x\sigma^{x}_{4}$ on the vortex-free ground state generate the final state, which is also the vortex-free ground state.
Thus, these perturbation processes are allowed within the ground state sector.
On the other hand, $\mathcal{H}_{\Gamma'}$ and $\mathcal{H}_{J_{H}}$ do not generate perturbation processes which lead to the the armchair-shaped  interaction satisfying the above condition.
Thus, we omit $\mathcal{H}_{\Gamma'}$ and $\mathcal{H}_{J_{H}}$ in $V'$ in this perturbative calculation.
Then, the third-order perturbation term is given by,
\begin{eqnarray}
\mathcal{H}^{(3)}_{\textrm{armchair}}&=&\Pi_{0}V^{\prime}G'_{0}(E)V^{\prime}G'_{0}(E)V^{\prime}\Pi_{0},\nn\\
&= &\frac{3{\Gamma}}{\Delta\Delta'}
 {\displaystyle \sum_{\begin{subarray}{c}   \alpha = x,y,z  \\ \beta \neq \alpha \\ \gamma \neq \beta, \alpha
 \\ \end{subarray}}
 \sum_{\begin{subarray}{c}  \langle kl\rangle_{\alpha} \\ \langle jk\rangle_{\beta} \\ \langle jl\rangle_{\gamma} \\ \end{subarray}}}   
 \Pi_0 h_\beta h_\gamma \sigma^{\beta}_{i}\sigma^{\gamma}_{j}\sigma^{\beta}_{k}\sigma^{\gamma}_{l}\Pi_0,\label{eq:armshaped}
\end{eqnarray}
where $\Delta'\sim0.23J$ is the energy gap of two visons for the configuration shown in Fig.~\ref{takahashi-supp2}(b) \cite{Kitaev2006}.
In the Majorana representation, Eq.~\eqref{eq:armshaped} is recast into,
\begin{eqnarray}
\mathcal{H}^{(3)}_{\textrm{armchair}}&=\Pi_0\Biggl[
{\displaystyle \sum_{\begin{subarray}{c} \langle kl\rangle_{\alpha} \\ \langle ik\rangle_{\beta}  \\ \langle jl\rangle_{\gamma} 
 \\ \end{subarray}}}
 (-\frac{3{\Gamma}h_{\beta}h_{\gamma}}{J^2})c_{i}c_{k}c_{l}c_{j}\hat{u}^{\beta}_{ik}\hat{u}^{\gamma}_{lj}\nn
\Biggr]\Pi_0. \label{eq:armshapedex}
\end{eqnarray}
This results in the armchair-shaped interaction of four neighboring Majorana fermions on the sites  $i$, $j$, $k$, $l$ shown in Fig.~\ref{takahashi-supp3}(e).

Finally, we consider the zigzag-shaped  interaction, which is generated by the third-order perturbation with respect to the Heisenberg term $\mathcal{H}_{J_H}$ and the Zeeman term $\mathcal{H}_Z$.
An example of the perturbation processes is shown in Fig.~\ref{takahashi-supp3}.
In this example of the third-order perturbations, the operations of 
$J_{H} \sigma^{x}_{1} \sigma^{x}_{2}$ and $h_{x}\sigma^{x}_{3}$ and $h_x\sigma^{x}_{4}$ on the vortex-free ground state generate the final state,
which is also the vortex-free ground state.
Thus, these perturbation processes are allowed within the ground state sector.
On the other hand, $\mathcal{H}_{\Gamma}$ and  $\mathcal{H}_{\Gamma'}$ do not contribute to the generation of the zigzag-shaped interaction,
and are omitted in the following calculations.
Then, the third-order perturbation term is given by,
\begin{eqnarray}
\mathcal{H}^{(3)}_{\textrm{zigzag}}&=&\Pi_{0}V^{\prime}G'_{0}(E)V^{\prime}G'_{0}(E)V^{\prime}\Pi_{0},\nn\\
&= &\frac{{3}}{\Delta\Delta''}
 {\displaystyle \sum_{\begin{subarray}{c}   \alpha = x,y,z  \\ \beta \neq \alpha
 \\ \end{subarray}}
 \sum_{\begin{subarray}{c}  \langle ij\rangle_{\alpha} \\ \langle ik\rangle_{\beta} \\ \langle jl\rangle_{\beta} \\ \end{subarray}}}   
 \Pi_0 {h_\beta}^{2} \sigma^{\beta}_{k}J_{H}\sigma^{\beta}_{i}\sigma^{\beta}_{j}\sigma^{\beta}_{l}\Pi_0,\label{eq:zigzagshaped}
\end{eqnarray}
where $\Delta''$ represents the energy gap of four visons as illustrated in Fig.~\ref{takahashi-supp3}(b). In the Majorana representation, Eq.~\eqref{eq:zigzagshaped} is recast into,
\begin{eqnarray}
\mathcal{H}^{(3)}_{\textrm{zigzag}}&=\Pi_0\Biggl[
{\displaystyle \sum_{\begin{subarray}{c} \langle ij\rangle_{\alpha} \\ \langle ik\rangle_{\beta}  \\\langle jl\rangle_{\beta} 
 \\ \end{subarray}}}
 (-\frac{3J_{H}{h_{\beta}}^{2}}{\Delta\Delta''})  c_{k}c_{i}c_{j}c_{l}\hat{u}^{\beta}_{ki}\hat{u}^{\beta}_{jl}\nn
\Biggr]\Pi_0. \label{eq:zigzagshapedex}
\end{eqnarray}
This amounts to the zigzag-shaped interaction of four neighboring Majorana fermions on the sites $i$, $j$, $k$, $l$ shown in Fig.~\ref{takahashi-supp3}(e).

We, here, stress again that up to the third-order perturbations, the above analysis exhausts all four-body interactions among neighboring Majorana fermions which are generated by the combination of applied magnetic fields and the Heisenberg or symmetric off-diagonal exchange interactions. 

We also take into account the renormalization of  the nearest-neighbor and next-nearest-neighbor hopping amplitudes of itinerant Majorana fermions due to
non-Kitaev interactions and magnetic fields. 
Then, finally, we obtain the effective Hamiltonian for itinerant Majorana fermions up to the third order perturbation,
\begin{eqnarray}
\mathcal{H}_{\rm eff}
&=&\sum_{a=1,2,3}t_a\sum_{\langle jk\rangle_a}ic_jc_k
+\sum_{a=4,5,6}t_a\sum_{\langle\!\langle jk\rangle\!\rangle_a}ic_jc_k\nonumber\\
&&+g\left[\sum_{\textrm{Y}}c_jc_kc_lc_m+\sum_{\textrm{Y'}}c_jc_kc_lc_m\right] \nonumber\\
&&+ \mathcal{H}^{(3)}_{\rm armchair}+ \mathcal{H}^{(3)}_{\rm zigzag},
\end{eqnarray}
where the definitions of the indices of the hopping amplitudes, $a=1\sim 6$, are similar to 
those shown in Fig.~\ref{SMfig1}(b), and $\langle\!\langle jk\rangle\!\rangle_a$ means an $a$-bond connecting
next-nearest-neighbor sites $j$ and $k$, and,
\begin{align}
t_1=J-\frac{2{J_H}^2}{\Delta^{\prime}}-\frac{12\Gamma^3}{{\Delta^{\prime}}^2}+\frac{2{h_{x}}^2}{\Delta},\label{eq:j1}
\end{align}
\begin{align}
t_2=J-\frac{2{J_H}^2}{\Delta^{\prime}}-\frac{12\Gamma^3}{{\Delta^{\prime}}^2}+\frac{2{h_{y}}^2}{\Delta},\label{eq:j2}
\end{align}
\begin{align}
t_3=J-\frac{2{J_H}^2}{\Delta^{\prime}}-\frac{12\Gamma^3}{{\Delta^{\prime}}^2}+\frac{2{h_{z}}^2}{\Delta},\label{eq:j3}
\end{align}
\begin{align}
t_4={\kappa}'-\frac{2\Gamma'(h_y+h_z)}{\Delta}+\frac{6{{\Gamma}'}^2h_x}{{\Delta}^2}+\frac{6\Gamma\Gamma^{\prime}}{\Delta\Delta^{\prime}}(2h_x+h_y+h_z),
\end{align}
\begin{align}
t_5={\kappa}'-\frac{2\Gamma'(h_z+h_x)}{\Delta}+\frac{6{{\Gamma}'}^2h_y}{{\Delta}^2}+\frac{6\Gamma\Gamma^{\prime}}{\Delta\Delta^{\prime}}(h_x+2h_y+h_z),
\end{align}
\begin{align}
t_6={\kappa}'-\frac{2\Gamma'(h_x+h_y)}{\Delta}+\frac{6{{\Gamma}'}^2h_z}{{\Delta}^2}+\frac{6\Gamma\Gamma^{\prime}}{\Delta\Delta^{\prime}}(h_x+h_y+2h_z),
\end{align}
\begin{align}
g =-\kappa'-\frac{6{{\Gamma}'}^2(h_x+h_y+h_z)}{{\Delta}^2},
\end{align}
with $\kappa'\equiv6h_xh_yh_z/{\Delta}^2$.
In the main text, we put $t_1=t_2=t_3=t$, and $t_4=t_5=t_6=\kappa$ to highlight spontaneous rotational-symmetry breaking due to four-Majorana interactions.
It is worth mentioning that, as seen in Eqs.(\ref{eq:j1})-(\ref{eq:j3}), the normalization due to the Heisenberg interaction 
and the $\Gamma$ term with $\Gamma > 0$,
reduces the nearest-neighbor hopping amplitudes in the case of the ferromagnetic Kitaev interaction, $J > 0$.
This remarkable feature implies that for real candidate materials such as $\alpha$-RuCl$_3$, where the magnitudes of $\Gamma$ and $J$ are in the same order,
the band width of itinerant Majorana fermions is substantially reduced, and thus, the systems may be in strongly correlated regions with 
$g \sim t$ in Eq.(5) of the main text, for which the nematic transition occurs.

\end{appendix}

\bibliography{paper}

\end{document}